\documentclass[12pt]{iopart_mod}

\usepackage{amsmath}
\usepackage{amsfonts}
\usepackage{mathrsfs,cite}
\usepackage{graphicx,color} 
\usepackage{epsfig}

\newcommand{\beqs}{\begin{equation*}}
\newcommand{\beq}{\begin{equation}}

\newcommand{\eeqs}{\end{equation*}}
\newcommand{\eeq}{\end{equation}}

\newcommand{\beqas}{\begin{eqnarray*}}
\newcommand{\beqa}{\begin{eqnarray}}

\newcommand{\eeqas}{\end{eqnarray*}}
\newcommand{\eeqa}{\end{eqnarray}}

\newcommand{\eq}[2]{\begin{equation} #1 \label{#2} \end{equation}}

\newcommand{\al}{\alpha}

\newcommand{\de}{\delta}

\newcommand{\ka}{\kappa}
\newcommand{\la}{\lambda}

\newcommand{\blist}{\begin{itemize}}
\newcommand{\elist}{\end{itemize}}

\DeclareMathOperator{\extdm}{d}
\newcommand{\extd}{\extdm \!}

\providecommand{\href}[2]{#2}

\newcommand{\bea}{\begin{eqnarray}}
\newcommand{\eea}{\end{eqnarray}}
\newcommand{\ba}{\begin{array}}
\newcommand{\ea}{\end{array}}
\newcommand{\ee}{\end{equation}}

\begin{document}

\title[Semi-classical unitarity in 3-dimensional non-principal higher-spin gravity]{Semi-classical unitarity in 3-dimensional higher-spin gravity for non-principal embeddings}
\author{H. Afshar, M. Gary, D. Grumiller, R. Rashkov and M. Riegler}

\address{  Institute for Theoretical Physics,
           Vienna University of Technology,\\
           Wiedner Hauptstr.~8-10/136,
           Vienna, A-1040, Austria} 

\eads{\mailto{afshar@hep.itp.tuwien.ac.at}, \mailto{mgary@hep.itp.tuwien.ac.at}, \mailto{grumil@hep.itp.tuwien.ac.at}, \mailto{rash@hep.itp.tuwien.ac.at}, \mailto{rieglerm@hep.itp.tuwien.ac.at}}

\begin{abstract}
Higher-spin gravity in three dimensions is efficiently formulated as a Chern--Simons gauge-theory, typically with gauge algebra $sl(N)\oplus sl(N)$. The classical and quantum properties of the higher-spin theory depend crucially on the embedding into the full gauge algebra of the $sl(2)\oplus sl(2)$ factor associated with gravity. It has been argued previously that non-principal embeddings do not allow for a semi-classical limit (large values of the central charge) consistent with unitarity. In this work we show that it is possible to circumvent these conclusions. Based upon the Feigin--Semikhatov generalization of the Polyakov--Bershadsky algebra, we construct infinite families of unitary higher-spin gravity theories at certain rational values of the Chern--Simons level that allow arbitrarily large values of the central charge up to $c = N/4 - 1/8 - {\cal O}(1/N)$, thereby confirming a recent speculation by us \cite{Afshar:2012nk}.
\end{abstract}

%\tableofcontents

%%%%%%%%%%%%%%%%%%%%%%%%%%%%%%%%%%%%%%%%%%%%%%%%%%%%%%%%%%%%%%%%%%%%%%%%%%%%%%%

%%%%%%%%%%%%%%%%%%%%%%%%%%%%%%%%%%%%%%%%%%%%%%%%%%%%%%%%%%%%%%%%%%%%%%%%%%%%%%%

\section{Introduction}

Quantum gravity is beset with conceptual and technical problems. An astucious strategy for progress is to eliminate as much of the technical issues as reasonably achievable. If successfully implemented, this strategy then allows one to tackle the interesting conceptual issues. This philosophy is behind many of the approaches to 3-dimensional quantum gravity that emerged in the past five years (see \cite{Witten:2007kt,Maloney:2007ud, Li:2008dq,Carlip:2008jk,Grumiller:2008qz,Giribet:2008bw,Bergshoeff:2009hq,Henneaux:2009pw,Liu:2009bk,Maloney:2009ck,Deser:2009hb,Bergshoeff:2009aq,Becker:2009mk,Skenderis:2009nt,Grumiller:2009mw,Grumiller:2009sn,Percacci:2010yk,Alishahiha:2010bw,Paulos:2010ke,Henneaux:2010xg,Campoleoni:2010zq,Grumiller:2010tj,Gaberdiel:2010pz,Castro:2010ce,Gutperle:2011kf,Chang:2011mz,Castro:2011zq,Castro:2012bc,Gaberdiel:2012uj} for an incomplete selection of references). Indeed, considering models of 3-dimensional gravity instead of 4- or higher-dimensional ones leads to drastic simplifications at the technical level, for instance the vanishing of the Weyl tensor. And yet, the baby has not been thrown out with the bathwater, since 3-dimensional gravity models exhibit basically all the features that we care about in quantum gravity: they can produce black hole solutions, local graviton excitations, allow for different asymptotic backgrounds and usually are susceptible to holographic descriptions. As an added bonus, whenever there exists a dual description in terms of a quantum field theory, that theory is usually formulated in 2 dimensions, and there is a lot of technical control over 2-dimensional quantum field theories in general and conformal field theories (CFTs) in particular.

Despite respectable progress in the past five years, there is still no family of quantum gravity models with the following properties:
\begin{itemize}
 \item All models are topological, in the sense that there are no local degrees of freedom.
 \item All models are unitary.
 \item Within the family, both ${\cal O}(1)$ and large values of the central charge are possible.
\end{itemize}
The first property is mostly a technical simplification, following the time-honored strategy to search for the keys first under the lamp-post; by accident or providence they might be there, after all. While it cannot be ruled out that dropping this assumption leads to a satisfactory family of quantum gravity models (for instance in the context of minimal model holography \cite{Gaberdiel:2010pz,Gaberdiel:2012uj}), it is much harder to study such models on the gravity side. Note that Einstein gravity and higher-spin gravity obey the first property. There are certainly physics applications where the second property has to be dropped --- for instance, in the context of the AdS/logarithmic CFT correspondence \cite{Grumiller:2008qz,Henneaux:2009pw,Maloney:2009ck,Skenderis:2009nt,Grumiller:2010tj} --- but in the context of quantum gravity toy models the second property seems like a condicio sine qua non. The third property is crucial for physics reasons. Some of the interesting conceptual questions that emerge in quantum gravity have to do with the discrepancy between naive semi-classical expectations, like information loss, and their quantum resolutions. To address these issues it is necessary to be able to discuss both the semi-classical limit (large central charge) and the quantum limit (central charge of order of unity). If only infinitely large central charges are allowed, the model may miss subtle quantum corrections and most likely will not resolve the semi-classical puzzles. If only ${\cal O}(1)$ values of the central charge are allowed, then the theory is `ultra-quantum' and typically does not permit any interpretation in terms of geometric entities, like black hole horizons. A snapshot of part of the landscape of 3-dimensional quantum gravity models with the first two properties is contained in \cite{Castro:2011zq}. None of the models studied there has the third property.

The absence of models with all the properties listed above motivates us to search for large classes of novel models. In this paper we consider a special class of 3-dimensional quantum gravity models, namely higher-spin gravity. (For general aspects of higher spin gravity see for instance \cite{Vasiliev:2012vf,Maldacena:2012sf,Giombi:2012ms} and references therein. For a review with focus on 3-dimensional black holes see \cite{Ammon:2012wc}.) We use the gauge-theoretic formulation of 3-dimensional higher-spin gravity, whose bulk action can be written as the difference of two $SL(N)$ Chern--Simons actions:
\eq{
S_{\rm bulk} = \frac{k_{\textrm{\tiny CS}}}{4\pi}\,\int_{\cal M}\!\!\! \tr\big[\textrm{CS}(A)-\textrm{CS}(\bar A)\big]
}{eq:hs1}
We denote the Chern--Simons level by $k_{\textrm{\tiny CS}}$. The Chern--Simons 3-form is given by
\eq{
\textrm{CS}(A)= A\wedge \extd A + \tfrac23\,A\wedge A \wedge A
}{eq:hs2}
and similarly for CS$(\bar A)$. For our purposes the manifold ${\cal M}$ is required to have cylindrical or solid torus topology, and the boundary cylinder/torus is the asymptotic boundary, so that the dual field theory (if it exists) lives either on a cylinder or on a torus.

So far, most of the work in 3-dimensional higher spin gravity employed the principal embedding of $sl(2)$ into $sl(N)$. Our focus will be on non-principal embeddings, partly because they outclass the principal embedding by sheer number, so one may hope to find large classes of new interesting models. We shall explain in more detail as we go along what these non-principal embeddings are, and why they are of interest. For the moment it is sufficient to highlight a property of non-principal embeddings that casts a shadow of gloom on their utility as quantum gravity models \cite{Castro:2012bc}: All non-principal embeddings have a singlet factor that leads to a Kac--Moody algebra as part of the asymptotic symmetry algebra,
\eq{
[J_n,\,J_m] = \kappa\,n\,\de_{n+m,\,0} + \dots
}{eq:hsg1}
where the ellipsis refers to possible non-abelian terms.
Unitarity requires non-negative level $\kappa$, since otherwise negative norm states appear in the spectrum of states. On the other hand, all embeddings by construction have an $sl(2)$ factor that leads to a Virasoro algebra as part of the asymptotic symmetry algebra,
\eq{
[L_n,\,L_m] = (n-m)\,L_{n+m} + \frac{c}{12}\,n(n^2-1)\,\de_{n+m,\,0}\,.
}{eq:hsg2}
Unitarity requires non-negative central charge $c$, for the same reasons as above. The key observation of  \cite{Castro:2012bc} was that $\kappa$ and $c$ are not independent quantities, and that for large values of the central charge its sign is always opposite to the sign of $\kappa$. 
\eq{
\textrm{sign}(c) = - \textrm{sign}(\kappa)\qquad \textrm{if}\; |c|\to\infty
}{eq:hsg3}
This argument then leads to the conclusion that semi-classical\footnote{By `semi-classical' we always mean the limit of large central charge, as explained in section \ref{se:2.1}.} unitarity is impossible for non-principal embeddings.

In the present paper we circumvent this conclusion and find an infinite family of non-principal embeddings where the central charge $c$ and the level $\kappa$ have the same sign, and yet we can make the central charge macroscopically large. While we are not able to make it infinite --- after all this would contradict the result \eqref{eq:hsg3} --- we discover that the value of the central charge where unitarity is possible is bounded by $N$, which we assume to be arbitrarily large but finite.
\eq{
c \leq \frac{N}{4} - \frac18 - {\cal O}(1/N)
}{eq:hsg4}
When approaching the bound \eqref{eq:hsg4} arbitrarily large values of the central charge are possible if we allow for arbitrarily large $N$. Moreover, depending on the level $k_{\textrm{\tiny CS}}$ certain rational values $c={\cal O}(1)$ are possible as well, thus obeying the third property above.

This paper is organized as follows. 
In section \ref{se:2} we review aspects of low spin holography (spin 2 and 3) to set the stage. We also give new results on spin 4 holography. 
In section \ref{se:3} we provide motivations to study specifically non-principal embeddings and present our main results that establish semi-classical unitarity for $W_N^{(2)}$ gravity. We derive the bound \eqref{eq:hsg4} announced above.
In section \ref{se:4} we conclude with an outlook to possible generalizations and open issues.

\section{Low spin holography}\label{se:2}

In this section we review salient aspects of low spin holography.
In section \ref{se:2.1} we recall some problematic issues with spin~2 holography.
In section \ref{se:2.2} we review recent results on spin~3 holography in the non-principal embedding, both for AdS and non-AdS backgrounds.
In section \ref{se:2.3} we present new results on spin~4 holography that generalize the results of section \ref{se:2.2}.

\subsection{Spin 2}\label{se:2.1}

Einstein gravity with negative cosmological constant is an interesting toy model for quantum gravity \cite{Witten:2007kt}: It is locally trivial \cite{Deser:1984tn,Deser:1984dr}, has BTZ black hole solutions \cite{Banados:1992wn}, and boundary graviton excitations that fall into representations of two copies of the Virasoro algebra \cite{Brown:1986nw}. The asymptotic symmetry algebra (ASA) is the conformal algebra with central charges determined by the Newton constant $G_N$ and the AdS radius $\ell$, $c=\bar c=3\ell/(2G_N)$ \cite{Brown:1986nw}.
An interesting attempt to identify the CFT dual of Einstein gravity \cite{Witten:2007kt} led to a renewed interest in this subject, and culminated in an identification between the (tricritical) Ising model and Einstein gravity for specific values of the Newton constant, $G_N=3\ell$ ($G_N=15\ell/7$) \cite{Castro:2011zq}. However, so far these are the only values of Newton's constant where a CFT dual for Einstein gravity could be identified, and both lead to a central charge smaller than unity. Therefore, there is currently no way to take a semi-classical limit of Einstein gravity on the CFT side.

Perhaps these difficulties might have been anticipated; after all, the theory contains black holes, and it is not quite clear how all the microstates responsible for the huge BTZ black hole entropy can be accommodated within pure Einstein gravity at small values of the Newton constant. Maybe we should not be surprised that only a small number of states (or small values of the central charge) can arise in an AdS/CFT context that involves exclusively Einstein's theory on the gravity side. As we argued in the introduction, it is not quite clear what we can learn physics-wise from a quantum gravity toy model whose semi-classical limit is unknown or inaccessible.

There seem to be two ways out. Either we get rid of the BTZ black holes, or we introduce more states in the gravity theory. The first option was pursued recently \cite{warped}. Namely, imposing Lobachevsky boundary conditions in conformal Chern--Simons gravity leads to a theory that has no black hole solutions, with an ASA that consists of a $\hat u(1)$ current algebra and a Virasoro algebra, just like in \eqref{eq:hsg1}, \eqref{eq:hsg2}. Moreover, the central charge $c$ has the same sign as the level $\kappa$, $c=24\kappa$, which is encouraging. However, the 1-loop calculation remains inconclusive, and no specific field theory dual has been identified yet. The second, potentially more interesting option requires the introduction of field degrees of freedom besides the metric, like in string theory. Given that we want to maintain the first property in the introduction, the absence of local physical degrees of freedom, we need to introduce a (bulk) gauge degree of freedom for each additional field degree of freedom. A first step in this direction is the consideration of holographic duals for conformal Chern--Simons gravity \cite{Afshar:2011yh}, %,Afshar:2011qw}, 
but this introduces merely one additional tensorial field degree of freedom, whereas the problem with BTZ black holes mentioned above indicates that we need to be able to add an arbitrary number of field degrees of freedom. Thus, it is natural to consider higher-spin theories, since they allow an arbitrary number of field degrees of freedom, but nevertheless remain topological.

\subsection{Spin 3}\label{se:2.2}

Principally embedded higher spin gravity in an AdS/CFT context was considered first by Henneaux and Rey \cite{Henneaux:2010xg}, and independently by Campoleoni, Fredenhagen, Pfenninger and Theisen \cite{Campoleoni:2010zq}. For several purposes it is of interest to consider also non-principally embedded higher spin gravity, in particular since the number of non-principal embeddings grows with $N$. Another reason to consider non-principal embeddings arises when the desired background configuration requires the presence of a singlet \cite{Gary:2012ms}, which exists only for non-principal embeddings.\footnote{As discussed in that paper, a singlet is needed for most non-AdS holography constructions like Lifshitz holography \cite{Kachru:2008yh,Ross:2011gu}, Schr\"odinger or null warped holography \cite{Son:2008ye,Balasubramanian:2008dm,Adams:2008wt}, or Lobachevsky holography \cite{warped}.} The principal embedding has the property that the spins arising in the decomposition of $sl(N)$ into $sl(2)$ representations are the integers $2,\,3,\,\dots,\,N$, which justifies the name ``spin $N$ gravity''. To simplify the language, we shall also refer to spin $N$ gravity even for the non-principal embeddings, where the highest spin is always lower than $N$. In other words, spin $N$ gravity is $sl(N)\oplus sl(N)$ Chern--Simons theory \eqref{eq:hs1} with some specific embedding of $sl(2)$ into $sl(N)$ and suitable boundary conditions for the connections. Finding the latter can be an art, but there are useful guidelines available, see for instance \cite{Afshar:2012nk}.

A simple example is provided by non-principally embedded spin 3 gravity with Lobachevsky boundary conditions \cite{Afshar:2012nk} (see also \cite{Riegler:2012}). Dropping all trivial fluctuations (those that do not contribute to the canonical boundary charges) the connections read
\begin{subequations}\label{AdSxR:BCs1}
		\begin{align}
A_t &=0 \qquad\qquad \bar{A}_t=\sqrt{3}\,S \qquad\qquad A_\rho = L_0 \qquad\qquad \bar{A}_\rho=-L_0 \\ 
A_\varphi &= -\frac{1}{4}\,L_1\,e^\rho + \frac{2\pi}{k_{\textrm{\tiny CS}}}\,\big(\mathcal{J}(\varphi)S+\mathcal{G}^{\pm}(\varphi)\psi^{\pm}_{-\frac{1}{2}}\,e^{-\rho/2}+\mathcal{L}(\varphi)L_{-1}\,e^{-\rho}\big)\\
\bar{A}_\varphi&=-L_{-1}\,e^\rho + \frac{2\pi}{k_{\textrm{\tiny CS}}}\,\bar{\mathcal{J}}(\varphi)S 
		\end{align}
\end{subequations}
Besides the $sl(2)$ generators $L_n$ and the singlet $S$, which are all present already in background quantities, the two doublets $\psi^\pm_{\pm 1/2}$ appear in the state-dependent fluctuations parametrized by various free functions of the angular coordinate $\varphi$. Their Fourier-components essentially constitute the generators of the ASA.

The boundary conditions \eqref{AdSxR:BCs1} lead to finite, integrable and conserved canonical charges that generate an ASA consisting of one copy of the Polyakov--Bershadsky algebra \cite{Polyakov:1989dm,Bershadsky:1990bg} and a $\hat u(1)$ current algebra.
Defining $k = -k_{\textrm{\tiny CS}}-3/2$ and denoting normal ordering by $::$, the ASA is given by 
	\begin{subequations}\label{AdSxR:W32AlgebraQuantumBershadsky}
		\begin{align}
			&[J_n,\,J_m]=\kappa\,n\,\delta_{n+m,\,0} = [\bar J_n,\bar J_m] 
\end{align}
\begin{align}
			&[J_n,\,L_m]=nJ_{n+m}\\
			&[J_n,\,G_m^{\pm}]=\pm G_{m+n}^{\pm}\\
%\end{align}
%\begin{align}
			&[L_n,\,L_m]=(n-m)L_{m+n}+\frac{c}{12}\,n(n^2-1)\,\delta_{n+m,\,0}\\
			&[L_n,\,G_m^{\pm}]=\big(\frac{n}{2}-m\big)\,G_{n+m}^{\pm}\\
			&[G_n^{+},\,G_m^{-}]= \frac{\lambda}{2}\,\big(n^2-\frac{1}{4}\big)\,\delta_{n+m,\,0}\nonumber \\
&\qquad\qquad\quad -(k+3)L_{m+n}+\frac{3}{2}(k+1)(n-m)J_{m+n}
			+3\sum_{p\in\mathbb{Z}}:J_{m+n-p}J_p:
		\end{align}
	\end{subequations}
with the $\hat u(1)$ level 
\eq{
\kappa = \frac{2k+3}{3}
}{eq:hsg8}
the Virasoro central charge
\begin{equation}
c=25-\frac{24}{k+3}-6(k+3) \label{eq:c}
\end{equation}
and the central term in the $G^\pm$ commutator
\eq{
\la = (k + 1)(2k + 3)\,.
}{eq:hsg9}
Non-negativity of $c$ requires the level $k$ to lie in the interval $-\tfrac13\geq k\geq-\tfrac32$. These inequalities exclude the possibility of a unitary field theory dual in the semi-classical limit $|c|\to\infty$, as mentioned in the introduction.

Another obstruction to unitarity comes from the $G^\pm$ sector. It turns out that the two level $3/2$ descendants of the vacuum, $G^\pm_{-3/2}|0\rangle$, lead to a Gram matrix proportional to $\la$ with positive and negative Eigenvalue.
Therefore, generic values of the level $k$ lead to positive and negative norm states, which makes the theory non-unitary. The only exception arises if $\la$ vanishes, in which case the $G^\pm$ descendants become null states. Thus, the number of possible values of $k$ compatible with unitarity is reduced to two values, $k=-1$ and $k=-3/2$. The latter leads to a trivial ($c=0$) theory, the former to a rather simple one ($c=1$).

The unitarity analysis above applies also to AdS holography in the non-principal embedding, where the ASA consists of two copies of the Polyakov--Bershadsky algebra (see \cite{Ammon:2011nk,Campoleoni:2011hg}). In summary, the non-principal embedding of spin 3 gravity provides a fairly modest step towards semi-classicality as compared to Einstein gravity: The central charge no longer has to be smaller than one, but can be equal to one. To decide whether this trend continues and eventually allows macroscopically large values of the central charge we have to consider spin $N$ gravity with larger values of $N$. Our next step is to consider non-principal embeddings of spin 4 gravity.

\subsection{Spin 4}\label{se:2.3}

Spin 4 gravity is the lowest spin gravity model where several qualitatively different non-principal embeddings exist (see \cite{Tan:2011tj,Gary:2012ms} for explicit results). The 2-1-1 embedding has four singlets and thus can lead to non-abelian current algebras as subalgebras of the ASA; in addition it has four doublets. The 2-2 embedding has four spin 2 excitations and three singlets. Finally, the 3-1 embedding (or ``next-to-principal embedding'') is the simplest non-principal embedding where a spin higher than 2 arises. We focus here on the 2-1-1 and 3-1 embeddings, since they are the closest analogs of the spin 3 non-principal embedding. Our goal is to investigate whether there are more possibilities to obtain unitary models than in the Polyakov--Bershadsky case.

\paragraph{2-1-1 embedding}

\renewcommand{\L}{\ensuremath{\mathcal{L}}}
\newcommand{\J}{\ensuremath{\mathcal{J}}}
\newcommand{\G}{\ensuremath{\mathcal{G}}}

In analogy to the case of the non-principal embedding of $sl(2)\hookrightarrow sl(3)$, in the 2-1-1 embedding of $sl(4)$ the unbarred asymptotic symmetry algebra is independent of whether we consider AdS or Lobachevsky boundary conditions, while the barred sector will either be identical to the unbarred sector or simply an $\widehat{su}(2)\oplus\hat{u}(1)$ current algebra, respectively. Therefore, in what follows, we will only consider the unbarred sector. Once again, dropping all trivial fluctuations, the connection is given by
\begin{subequations}
\begin{align}
A_t &= 0 \qquad\qquad A_\rho = L_0 \\
A_\phi &= L_1e^\rho + \frac{2\pi}{k_{\textrm{\tiny CS}}}\,\big(\J(\varphi)S %\nonumber\\ & \qquad\qquad\qquad 
+ \!\!\!\!\!\sum_{a=-1,\,0,\,1}\!\!\! \J^{a}(\varphi)S^{a}+\!\sum_{a,b=\pm}\! \G^{ab}(\varphi)\psi^{ab}_{-\frac{1}{2}}e^{-\rho/2} +\L(\varphi)L_{-1}e^{-\rho}\big)\,.
\end{align}
\end{subequations}
The generators $S$, $S^a$, $\psi^{ab}_n$ and $L_n$ refer to the $\hat u(1)$ singlet, $\widehat{su}(2)$ singlet, four doublets and the gravity triplet, respectively.
These boundary conditions lead to finite, integrable, and conserved canonical charges generating an ASA whose non-vanishing commutators are given by\footnote{It is also possible to consider a looser set of boundary conditions. Such boundary conditions lead to a slightly larger set of finite, integrable, conserved charges.} %However, the unitary representations of the asymptotic symmetry algebras seem likely to coincide, as the enlargement is purely in the $\G$ sector.}
\begin{subequations}
\begin{align}
  &[J_n,\,J_m]=\kappa\, n\,\delta_{n+m,\,0}\\
  &[J_n,\,L_m]=nJ_{n+m}\\
  &[J_n,\,G^{a\pm}_m]=\pm G^{a\pm}_{n+m}\\
  &[J^{a}_n,\,J^{b}_m]=(a-b)J_{m+n}^{a+b} + \kappa_2\,(1-3a^2)\,n\,\delta_{a+b,\,0}\delta_{n+m,0\,} \label{eq:13d} \\ 
  &[J^{a}_n,\,L_m]=nJ^{a}_{n+m}\\
  &[J^\pm_n,\,G^{\mp a}_m]=\pm G^{\pm a}_{n+m} = -2\,[J^0_n,\,G^{\pm a}_m] \\
%\end{align}
%\begin{align}
  &[L_n,L_m]=(n-m)L_{n+m}+\frac{c}{12}\,n(n^2-1)\,\delta_{m+n,\,0}\\
  &[L_n,G^{ab}_m]=\big(\frac{n}{2}-m\big)G^{ab}_{n+m} \\
%\end{align}
%\begin{align}
  &[G^{\pm\pm}_n,\,G^{\pm\mp}_m]=-k_{\textrm{\tiny CS}}(m-n)J^{\pm}_{m+n}\mp\sum_{p\in\mathbb{Z}}:\left(J_{m+n-p}J_p^{\pm}+J_{m+n-p}^{\pm}J_p\right): 
\end{align}
\begin{align}
  &[G^{++}_n,\,G^{--}_m]=\lambda\,\big(n^2-\frac{1}{4}\big)\,\delta_{m+n,\,0}  \nonumber \\
  &\qquad+(k_{\textrm{\tiny CS}}-2)L_{n+m}+(k_{\textrm{\tiny CS}}+1)(m-n)J_{m+n}-k_{\textrm{\tiny CS}}(m-n)J_{m+n}^{0} \nonumber \\
      +\sum_{p\in\mathbb{Z}} &:\Big(\tfrac{3}{2}J_{m+n-p}J_{p} 
- J_{m+n-p}J_{p}^{0} - J_{m+n-p}^{0}J_{p} + \sum_{a+b=0}(2-3a^2)J_{m+n-p}^aJ_p^b\Big):
\end{align}
\end{subequations}
with $\hat{u}(1)$ level
\begin{equation}
\kappa = -k_{\textrm{\tiny CS}}\, ,
\end{equation}
$\widehat{su}(2)$ central extension\footnote{The $\widehat{su}(2)$ central term in \eqref{eq:13d} can be brought into the standard form by a change of basis of the $su(2)$ generators $J^a$.}
\begin{equation}
\kappa_2 = -\frac{k_{\textrm{\tiny CS}}+1}{2}\, ,
\end{equation}
the Virasoro central charge
\begin{equation}
c = \frac{3(k_{\textrm{\tiny CS}}+2)(2k_{\textrm{\tiny CS}}+1)}{k_{\textrm{\tiny CS}}-2}\, ,
\end{equation}
and the central term in the $G^{ab}$ commutator
\begin{equation}
\lambda = k_{\textrm{\tiny CS}}(k_{\textrm{\tiny CS}}+1)\, .
\end{equation}

Searching for unitary representations of the ASA, non-negativity of the Virasoro central charge $c$ forces the level to lie within the region $-2\leq k_{\textrm{\tiny CS}} \leq-\frac{1}{2}$. Once again the strictest requirement comes from the fact that the $G^{++}_{-\frac{3}{2}}$ and $G^{--}_{-\frac{3}{2}}$ descendants of the vacuum have opposite norm and therefore must be null, forcing $k_{\textrm{\tiny CS}}=-1$. At the one unitary value of $k_{\textrm{\tiny CS}}$, the $\widehat{su}(2)$ level also vanishes, leaving just the simple theory of a single $\hat{u}(1)$ current algebra with Virasoro central charge $c=1$.

\paragraph{3-1 embedding --- $W_4^{(2)}$ gravity} This case is quite analog to the spin 3 non-principal embedding, so we restrict ourselves to presenting the main results. We denote the generators corresponding to the singlet, gravity triplet, two triplets and spin 3 generators by $S$, $L_n$, $\psi^\pm_n$ and $W_n$, respectively. 
Feeding the boundary conditions 
\begin{subequations}\label{eq:angelinajolie}
		\begin{align}
A_t &=0 \qquad\qquad \bar{A}_t=\sqrt{2/\tr(S^2)}\,S \qquad\qquad A_\rho = L_0 \qquad\qquad \bar{A}_\rho=-L_0 \\ 
A_\varphi &= 1/[4\tr(L_1 L_{-1})]\,L_1\,e^\rho + \frac{2\pi}{k_{\textrm{\tiny CS}}}\,\big( \mathcal{J}(\varphi)S+\mathcal{G}^{\pm}(\varphi)\psi^{\pm}_{-1} \,e^{-\rho} \nonumber \\ &\qquad\qquad\qquad\qquad\qquad\qquad 
+\mathcal{L}(\varphi)L_{-1}\,e^{-\rho}+\mathcal{W}(\varphi)W_{-2}\,e^{-2\rho} \big)\\
\bar{A}_\varphi&=-L_{-1}\,e^\rho + \frac{2\pi}{k_{\textrm{\tiny CS}}}\,\bar{\mathcal{J}}(\varphi)S 
		\end{align}
\end{subequations}
into the algorithm described in \cite{Afshar:2012nk} eventually leads to the following quantum ASA
\begin{subequations}\label{eq:31}
\begin{align}
& [J_n,\,J_m] = \kappa\,n\,\de_{n+m,\,0} \\
& [L_n,\,L_m]=(n-m)L_{m+n}+\frac{c}{12}\,n(n^2-1)\,\delta_{n+m,\,0}\\
& [G^+_n,\,G^-_m] = \la \,\tfrac12\,n(n^2-1)\,\de_{n+m,\,0} + \dots 
\end{align}
\end{subequations}
We have provided only commutators with non-trivial central terms. The ellipsis in the last commutator refers to terms that are known but irrelevant for our discussion of unitarity. A more complete set of commutators is provided in \eqref{eq:FS} below. Note that the commutators between the spin 3 generators, $[W_n,\,W_m]$, also contain a central term. However, for non-vanishing $\kappa$ this term is proportional to $c-1$, and thus conveys no new information. The subtraction of $1$ comes from the Sugawara shift of the Virasoro generators, i.e., the total Virasoro central charge consists of a `bare' part $c-1$ and a Sugawara part of $1$. The central term appearing in the spin 3 commutator is proportional to the bare central charge. For vanishing $\kappa$ there is no Sugawara shift and all central terms vanish, $\kappa=c=\lambda=0$.

We provide now the results for the central terms in the ASA \eqref{eq:31}.
Defining $k=\tfrac{3}{16}\, k_{\textrm{\tiny CS}}$ the $\hat u(1)$ level is given by
\eq{
\kappa = \frac{3}{4}\,k + 2
}{eq:hsg29}
the Virasoro central charge by
\eq{
c =- \frac{(3k+8)(8k+17)}{k+4}
}{eq:hsg30}
and the central term in the $G^\pm$ commutator by
\eq{
\la = (k+2)(2k+5)(3k+8)\,.
}{eq:hsg31}
Taking into account all unitarity constraints leads to two solutions for the level, $k=-\tfrac83$ leading to $\kappa=c=\lambda=0$, and $k=-\tfrac52$, leading to $\ka=\tfrac18$, $c=1$ and $\la=0$. Thus, there is no additional unitary solution as compared to section \ref{se:2.2}.

Nevertheless, it is encouraging that the strictest condition on unitarity, $\la=0$, has one additional solution as compared to the spin 3 case. We show in the next section that this trend continues and allows one to find unitary models for arbitrarily large values of the central charge, provided we make the spin high enough.

\section{Non-principal holography for arbitrary spin}\label{se:3}

In this section we derive our main results.
In section \ref{se:3.1} we describe our aim and how we intend to achieve it. 
In section \ref{se:3.2} we focus on $W_N^{(2)}$ gravity, based upon the Feigin--Semikhatov \cite{Feigin:2004wb} generalization of the Polyakov--Bershadsky algebra.
In section \ref{se:3.3} we discuss the necessary conditions for unitarity in the semi-classical limit.
In section \ref{se:3.4} we give a physical interpretation of our results.
In section \ref{se:3.5} we provide AdS and Lobachevsky boundary conditions that lead to the desired ASA.

\subsection{What to aim for}\label{se:3.1}

Non-principal embeddings are required for typical non-AdS holography applications, but they could also be useful for AdS holography. For instance, the higher spin cases in the principal embedding studied in \cite{Castro:2011zq} all have central charges of order of unity, and it is not clear if there are consistent examples of principally embedded higher spin gravity models for arbitrary values of the central charge. Perhaps some of the non-principal embeddings provide useful models with the desired properties. Before speculating about such applications, however, we have to deal with (non-)unitarity in the semi-classical limit.
We have seen above that the naive semi-classical limit, $|c|\to\infty$, always leads to non-unitarity (as observed first in \cite{Castro:2012bc}), but this does not rule out the possibility of more sophisticated semi-classical limits. We follow here a somewhat optimistic suggestion in our earlier paper \cite{Afshar:2012nk} and show in the rest of this section that the optimism was justified.

Namely, suppose that we are given an arbitrarily large value of the Virasoro central charge $c$ and are forced to find a non-principal embedding that leads to this central charge, but without violating unitarity. If we are able to succeed then we have basically met our goal described in the introduction --- namely, a family of topological models that are unitary and allow for ${\cal O}(1)$ as well as arbitrarily large values of the central charge. The only logical possibility to succeed is to allow for arbitrary values of the spin $N$, since for every given value of $N$ there will be a value of the central charge that is too large to be compatible with unitarity. 

Thus, what we are aiming for is an infinite family of spin $N$ gravity models in some non-principal embedding, where the central charge grows in some way with $N$ and compatibility with unitarity can be achieved.

\subsection{$W_N^{(2)}$ gravity}\label{se:3.2}

We consider the next-to-principal embedding of $sl(2)$ into $sl(N)$ [the $(N-1)$-1 embedding] and call the ensuing theory $W_N^{(2)}$ gravity. By `next-to-principal' we mean that there occur all integer spins up to (including) $N-1$, but not spin $N$. In addition, there is always a pair of spin $N/2$ fields and a singlet. The counting works, since $N^2-1 = (N-1)^2-1 + 2(N-1) + 1$. The simplest case, $N=3$, leads to the Polyakov--Bershadsky algebra \eqref{AdSxR:W32AlgebraQuantumBershadsky}, with a spin 2 field, two spin 3/2 fields and the spin 1 current associated with the singlet. The next-simplest case, $N=4$, leads to the $W_4^{(2)}$ algebra encountered in section \ref{se:2.3} for the 3-1 embedding, with a spin 2 and spin 3 field, two additional spin 2 fields and again the spin 1 current associated with the singlet.

Since we are interested in unitarity, we work directly on the field theory side for the time being and study the expected quantum ASA, but we do not bother with boundary conditions on the gravity side until section \ref{se:3.5}. The expected quantum ASA for AdS (Lobachevsky) holography consists of two copies (a $\hat u(1)$ current algebra and one copy) of the $W_N^{(2)}$ algebra, introduced by Feigin and Semikhatov \cite{Feigin:2004wb}. 

One of the abstract algebraic constructions of $W$-algebras associated with
certain affine algebras $\hat{\mathfrak{g}}$ is roughly as follows. Consider
the quantum algebra $\mathcal{U}_q\big(\mathfrak{g}\big)$ and its centralizer
$\mathcal{A}$ in the algebra of local operators. Then $\mathcal{A}$
is a vertex algebra associated with the symmetries of the conformal field theory 
associated with the corresponding root system and is called $W$-algebra 
\cite{Feigin:1989,Feigin:1990pn,Frenkel:2004jn}.
The focus in \cite{Feigin:2004wb} is on the sequence of algebras $W^{(2)}_N$ generated by two currents $G^\pm_N$ of dimension $\frac{N}{2}$. The leading term in their
operator product expansion starts with an $N^\text{th}$ order pole representing a central term. The most
familiar and simplest member of this sequence is the Polyakov--Bershadsky $W^{(2)}_3$ algebra.
 The authors present an explicit construction of the $W^{(2)}_N$ algebra as a centralizer 
of the screenings representing the nilpotent subalgebra of $\mathcal{U}_q\big(sl(N|1)\big)$. 
The construction allows them to obtain the explicit form of the central charges, which is crucial for our considerations. A little bit less abstract way to look at the algebra is as a Drinfeld--Sokolov
reduction of a non-principal embedding of the $sl(2)$ algebra in $sl(N)$ --- in this particular case as a principal embedding of $sl(2)$ in $sl(N-1)\subset sl(N)$.

We review now the most relevant aspects of the algebra $W_N^{(2)}$ \cite{Feigin:2004wb}, which depends on the parameter $N$ and a level $k$ [not to be confused with the Chern--Simons level in \eqref{eq:hs1}].
\begin{subequations}\label{eq:FS}
\begin{align}
& [J_n,\,J_m] = \kappa\,n\,\de_{n+m,\,0} \label{eq:FSa}\\
& [J_n,\,L_m]=nJ_{n+m}\\
& [J_n,\,G_m^{\pm}]=\pm G_{m+n}^{\pm}\\
& [L_n,\,L_m]=(n-m)L_{m+n}+\frac{c}{12}\,n(n^2-1)\,\delta_{n+m,\,0}\\
& [L_n,\,G_m^\pm]= \big(n(\tfrac N2-1)-m\big)\,G_{n+m}^\pm\\
& [G^+_n,\,G^-_m] = \la \,f(n)\,\de_{n+m,\,0} + \dots \\
& [W_n^l,\,\textrm{anything}] = \dots
\end{align}
\end{subequations}
The commutators with current algebra generators $J_n$, Virasoro generators $L_n$ and the pair of generators $G^\pm_n$ contain central terms whose precise form is crucial. By contrast, the commutators involving higher spin generators $W_n^l$ [with $l=3,4,\dots,(N-1)$] do not contain pivotal information, since their central charge is determined by the Virasoro central charge and is positive if the latter is bigger than unity.
The $\hat u(1)$ level is given by
\eq{
\kappa = \frac{N-1}{N}\,k + N - 2
}{eq:hsg5}
the Virasoro central charge by
\eq{
c =- \frac{\big((N+k)(N-1)-N\big)\big((N+k)(N-2)N-N^2+1\big)}{N+k}
}{eq:hsg6}
and the central term in the $G^\pm$ commutator by
\eq{
\la = \prod\limits_{m=1}^{N-1} \big(m(N+k-1)-1\big)
}{eq:hsg7}
while $f(n)\propto \prod_{j=-(N/2-1)}^{N/2-1}(n+j)$, so that the 2-point correlator takes the form $\langle G^+(z) G^-(0)\rangle = \la / z^N+\dots$ e.g., for the Polyakov--Bershadsky algebra \eqref{AdSxR:W32AlgebraQuantumBershadsky} $f(n)=\tfrac{n^2}{2}-\tfrac18$ and for the $W_4^{(2)}$ algebra \eqref{eq:31} $f(n)=\tfrac12\,n(n^2-1)$.
For $N=3$ ($N=4$) we recover from the formulas above the results from section \ref{se:2.2} (section \ref{se:2.3}, 3-1 embedding). 
Before proceeding let us note that for each value of the level $k$ there exists a dual value $\tilde k$ that leads to the same expression for the central charge \eqref{eq:hsg12}.
\eq{
\tilde k = \frac{N+1}{N-2}\,\frac{1}{N+k}-N
}{eq:hsg18}
The duality is involutive (as it must be), $\tilde{\tilde k}=k$. 

We assume from now on that $N$ is at least 3, and intend to take the limit of large (but finite) $N$. For this purpose it turns out to be useful to parametrize the level $k$ in terms of a constant $\alpha$, defined by
\eq{
k = - N + 1 + \frac{\alpha+1}{N-1}\,.
}{eq:hsg10}
The duality \eqref{eq:hsg18} acts on the parameter $\alpha$ as follows.
\eq{
\tilde \alpha = \frac{N\big(N(1-\alpha)+2\alpha-1\big)+1}{(N-2)(N+\alpha)}= 1-\alpha + {\cal O}(1/N)
}{eq:hsg19}
As $N$ tends to large positive values, the level $k$ approaches large negative values for $\alpha\sim{\cal O}(1)$, but in such a way that the sum of $N+k$ remains close to unity. Interestingly, the subleading term containing $\alpha$ in the definition \eqref{eq:hsg10} will play a crucial r\^ole in the discussion of unitarity.

\subsection{Semi-classical unitarity}\label{se:3.3}

\begin{figure}
\centering
 \epsfig{file=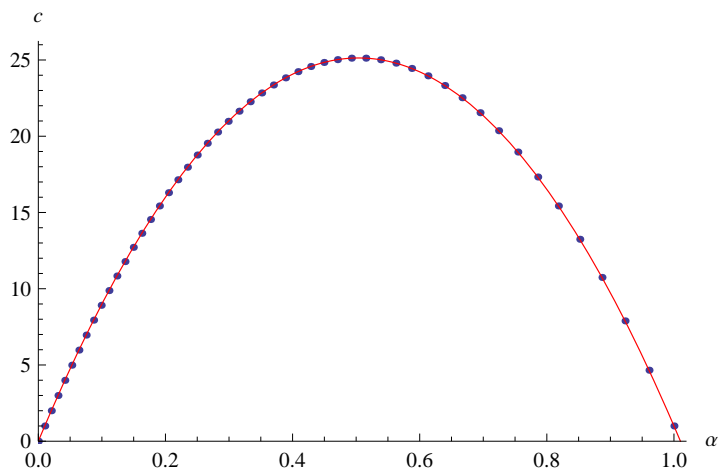,width=0.6\linewidth}
\caption{Virasoro central charge $c$ as function of the parameter $\alpha$ for $N=101$. Red solid curve allowed by positivity. Blue dots allowed by unitarity.}
\label{fig:1}
\end{figure}

Inserting the parametrization \eqref{eq:hsg10} into the result for the $\hat u(1)$ level \eqref{eq:hsg5} yields
\eq{
\kappa = \frac{\al}{N}
}{eq:hsg11}
Non-negativity of $\kappa$ imposes a first restriction on the parameter $\alpha$, 
\eq{
0 \leq \alpha
}{eq:hsg14}
A stronger constraint on $\alpha$ comes from the central charge \eqref{eq:hsg6}, which reads
\eq{
c = \al(1-\al)\,N + \alpha\,(\alpha^2+\alpha-1) - \sum\limits_{m=1}^{\infty} (1+\alpha)^2 (1-\alpha) \Big(-\frac{\alpha}{N}\Big)^m \,.
}{eq:hsg12}
Positivity of the central charge restricts $\al$ to the following interval.
\eq{
0\leq \alpha \leq \frac{N(N-1) + 1}{N(N-2)} 
}{eq:hsg13}
In the large $N$ limit $\alpha$ is then essentially restricted to the interval $[0,\,1]$. The inequalities \eqref{eq:hsg13} and \eqref{eq:hsg14} are compatible with each other. Therefore, non-negativity of the $\hat u(1)$ level is not at odds with non-negativity of the central charge. Moreover, the central charge \eqref{eq:hsg12} scales linearly with $N$, and thus can get arbitrarily large if we allow arbitrarily large spins, see Fig.~\ref{fig:1} where $c$ is plotted for $N=101$. These are encouraging results.

However, a huge obstacle remains. As in the $W_3^{(2)}$ case studied in \cite{Afshar:2012nk}, the $G^\pm$ sector generically leads to positive and negative norm states. The only exceptions arise again for vanishing central term $\lambda$ from \eqref{eq:hsg7}, since in this case the $G^\pm$ sector yields null states. Requiring $\lambda$ to vanish establishes a polynomial equation for $\alpha$ of degree $N-1$:
\eq{
\lambda =\prod\limits_{m=1}^{N-1} \Big(m\,\frac{\alpha+1}{N-1}-1\Big) \stackrel{!}{=} 0
}{eq:hsg15}
So there are $N-1$ solutions for $\alpha$ compatible with vanishing $\lambda$. 
\eq{
\lambda =0 \qquad\Leftrightarrow\qquad \alpha \in\Big\{ 0,\, \frac{1}{N-2}, \, \frac{2}{N-3},\, \dots,\, \frac{N-4}{3},\,\frac{N-3}{2},\,N-2 \Big\}
}{eq:hsg16}
All of them are real and non-negative, but not all of them obey the inequalities \eqref{eq:hsg13}.
Selecting those $\alpha$ from \eqref{eq:hsg16} that obey the inequalities \eqref{eq:hsg13} leads to $(N+1)/2$ solutions ($N/2$ solutions) for odd (even) $N$. In conclusion, we obtain the following lists of allowed rational values for $\alpha$:
\eq{
\alpha = \left\{\begin{array}{ll}
 0,\,\frac{1}{N-2}, \, \frac{2}{N-3},\, \frac{3}{N-4},\, \dots,\, \frac{N-7}{N+5},\, \frac{N-5}{N+3},\, \frac{N-3}{N+1},\, 1 &\quad N \;\textrm{odd}\\
 0,\,\frac{1}{N-2}, \, \frac{2}{N-3},\, \frac{3}{N-4},\, \dots,\, \frac{N-8}{N+6},\, \frac{N-6}{N+4},\, \frac{N-4}{N+2},\, 1-\frac{2}{N} &\quad N \;\textrm{even}\\
                \end{array}\right.
}{eq:hsg17}
For each $\alpha$ in these lists the $\hat u(1)$ level \eqref{eq:hsg11} and the Virasoro central charge \eqref{eq:hsg12} are non-negative, and the $G^\pm$ descendants of the vacuum are all null states. Moreover, for non-vanishing $c$ the inequality $c\geq 1$ holds. Since the central terms appearing in the $W^l$ part of the algebra are all proportional to $c-1$ (with positive proportionality constant) also these central terms are always non-negative. Thus, we can have unitary representations of the ASA \eqref{eq:FS} for precisely the values of $\alpha$ appearing in \eqref{eq:hsg17}.

\subsection{Physical discussion}\label{se:3.4}

We provide now a physical discussion in terms of the $\hat u(1)$ level $\kappa$ \eqref{eq:hsg11} and the Virasoro central charge $c$ \eqref{eq:hsg12}. 
As we continue forward in our list \eqref{eq:hsg17} the allowed values of $\alpha$ are given by
\eq{
\alpha = \frac{\hat N}{N-\hat N-1}\qquad\qquad \hat N\in\mathbb{N},\; N\geq 2\hat N+1\,.
}{eq:hsg20}
Defining $m:=N-2\hat N-1$ the central charge \eqref{eq:hsg6} simplifies to 
\eq{
c(\hat N,\,m) - 1 = \big(\hat N-1\big)\Big(1-\frac{\hat N(\hat N+1)}{(m+\hat N)(m+\hat N+1)}\Big) 
}{eq:hsg21}
Thus, we obtain for the central charge \eqref{eq:hsg21} the values of the $W_{\hat N}$-minimal models (see e.g.~\cite{Blumenhagen}), shifted by $1$ from the $\hat u(1)$ current algebra. For $\hat N=2$ and arbitrary $m$ the values of the Virasoro minimal models are obtained for the bare central charge $c-1$. For instance, spin 6 gravity with $\alpha=\tfrac23$ leads to $c-1=\tfrac12$, i.e., the bare central charge equals to the one of the Ising model.

For further discussion we divide the spectrum of allowed values of $\alpha$ \eqref{eq:hsg17} into three regimes, a quantum regime for small values of $\alpha$, a semi-classical regime for generic values of $\alpha$ in the interval $(0,\,1)$ and a ``dual'' quantum regime for values of $\alpha$ close to unity. By `dual' we refer to the fact that the duality \eqref{eq:hsg19} maps values of $\alpha$ close to zero to values of $\tilde \alpha$ close to unity. Note, however, that the duality does not necessarily map values of $\alpha$ at the beginning of the lists \eqref{eq:hsg17} to values of $\alpha$ at the end of the lists, but instead can lead to values of $\alpha$ not contained in these lists.

\paragraph{Quantum regime} We start with the strong coupling limit on the gravity side. This means we choose the Chern--Simons level $k_{\textrm{\tiny CS}}$ as small as possible, which implies that we should choose the smallest possible values of $\alpha$. The value $\alpha=0$ always is possible and trivial. It corresponds to $\kappa=c=0$, and the only state in the theory is the vacuum. The next possible value is $\al=1/(N-2)$, leading to a small value for the $\hat u(1)$ level, $\kappa \sim {\cal O}(1/N^2)$, and unity central charge, $c=1$. In this case not only the $G^\pm$ sector decouples, but also the $W^l$ sector. The dual field theory consists of a free boson, just like in the Polyakov--Bershadsky case \cite{Afshar:2012nk}. These first examples confirm the general belief that quantum gravity in the ultra-quantum limit might be dual to a very simple field theory.  In the quantum regime at large $N$ the central charge takes values close to integers, $c=0,\,1,\,2-{\cal O}(1/N^2),\,\dots$

\paragraph{Semi-classical regime} If we make the coupling constant weaker we eventually approach the semi-classical regime, in which the central charge becomes proportional to $N$ and has a quasi-continuous spacing. We are interested specifically in the region where the central charge is close to its maximum value, corresponding to $m\approx\tfrac23 N$ in \eqref{eq:hsg16} or $\alpha\approx\tfrac12$. We parametrize the integer $m$ as $m=\tfrac23(N-1+n_0)+n_1$, where $n_1$ is some varying integer with absolute value much smaller than $N$ and $n_0\in\{0,\,1,\,2\}$ is fixed so that $m$ is an integer [more precisely, $n_0=(2N+1)\;\textrm{mod}\;3$]. This parametrization yields
\eq{
\alpha = \frac12 + \frac{3(3n_1-2n_0)}{4N}\,.
}{eq:hsg23}
The central charge \eqref{eq:hsg6} simplifies to
\eq{
c(n_1) =  \frac N4 - \frac 18 - \frac{\delta c}{N} + {\cal O}(1/N^2)
}{eq:hsg22}
with $\delta c= 9 (9n_1^2 - 3(4n_0+1) n_1 + 4n_0^2+2n_0-1)/16$.
The maximum of the continuous curve $c$ (see e.g.~figure \ref{fig:1}) is never reached, but for large $N$ one can get arbitrarily close to it, provided $n_1$ is a small integer.
The result \eqref{eq:hsg22} proves the bound \eqref{eq:hsg4} announced in the introduction.

Other semi-classical regimes can be obtained for $m=\tfrac pq(N-1+n_0)+n_1$, with some co-prime integers $p$ and $q$ so that $\tfrac12<\tfrac pq<1$, a suitable choice for $n_0\in\{0,\,1,\,\dots,\,q-1\}$ and some varying integer $n_1$ with absolute value much smaller than $N$. In those regimes the central charge scales as 
\eq{
c(n_1;\,\tfrac pq) = (2-\tfrac qp) (\tfrac qp-1)\, N + \delta c + {\cal O}(1/N)\,.
}{eq:hsg24}
with $\delta c=(2 n_1+1)\tfrac{q^3}{p^3} - (3n_1 -2n_0+2)\tfrac{q^2}{p^2} - 3 n_0 \tfrac qp + 1 $.

\paragraph{Dual quantum regime} As we keep decreasing the coupling constant, we eventually surpass the maximal value of the central charge and come back to smaller values of $c$, until we finally reach again values of order of unity, $c\sim{\cal O}(1)$. This is the behavior expected from the duality \eqref{eq:hsg19}. However, the dual quantum regime differs in its detail from the the quantum regime discussed above. Even assuming that $N$ is large, we have to discriminate between even and odd values of $N$, as expected from the lists \eqref{eq:hsg17}. 
Let us start by assuming $N$ is odd. Then the allowed values for $\alpha$ are given by
\eq{
\alpha = \frac{N-2m-1}{N+2m-1}\qquad N \gg m\in\mathbb{N} \,.
}{eq:hsg25}
The central charge \eqref{eq:hsg6} simplifies to 
\eq{
c(m) = 1 + 4m - {\cal O}(1/N)\qquad\qquad N\;\textrm{odd}
}{eq:hsg26}
Explicitly, we obtain the values $c=1,\,5,\,9,\,\dots$ up to ${\cal O}(1/N)$ corrections.
For even values of $N$ we proceed in a similar way, choosing $\alpha=(N-2m-2)/(N+2m)$. The central charge \eqref{eq:hsg6} simplifies to 
\eq{
c(m) = 3 + 4m - {\cal O}(1/N)\qquad\qquad N\;\textrm{even}
}{eq:hsg27}
Explicitly, we obtain the values $c=3,\,7,\,11,\,\dots$ up to ${\cal O}(1/N)$ corrections.

Thus, in both cases the dual quantum regime leads to a central charge with level spacing of four to leading order in a $1/N$ expansion. This implies that in the large $N$ limit only a quarter of the quantum regime levels gets mapped to corresponding levels in the dual quantum regime. 
%Moreover, the fact that even at large $N$ we have to discriminate between odd and even $N$ suggests that a strict $N\to\infty$ limit might not exist, at least not for the dual quantum regime.

\subsection{AdS and Lobachevsky boundary conditions}\label{se:3.5}

We leave now the field theory side and discuss some aspects on the gravity side.
We start with Lobachevsky boundary conditions generalizing the ones displayed in \eqref{AdSxR:BCs1}.
\begin{subequations}\label{AdSxR:BCs2}
		\begin{align}
A_t &=0 \qquad\qquad \bar{A}_t=\sqrt{2/\tr(S^2)}\,S \qquad\qquad A_\rho = L_0 \qquad\qquad \bar{A}_\rho=-L_0 \\ 
A_\varphi &= 1/[4\tr(L_1L_{-1})]\,L_1\,e^\rho + \frac{2\pi}{k_{\textrm{\tiny CS}}}\,\big( \mathcal{J}(\varphi)S+\mathcal{G}^{\pm}(\varphi)\psi^{\pm}_{-N/2+1} \,e^{-(N/2-1)\rho}\nonumber \\
&\qquad\qquad\qquad\qquad\qquad + \mathcal{L}(\varphi)L_{-1}\,e^{-\rho}+\sum\limits_{m=3}^{N-1} \mathcal{W}_m (\varphi)W^m_{-m+1}\,e^{-(m-1)\rho} \big)\\
\bar{A}_\varphi&=-L_{-1}\,e^\rho + \frac{2\pi}{k_{\textrm{\tiny CS}}}\,\bar{\mathcal{J}}(\varphi)S 
		\end{align}
\end{subequations}
The generators $S, L_n, \psi^\pm_n$ and $W_n^m$ refer to the singlet, gravity triplet, the two $(N-1)$-plets and the spin $m$ generators, respectively. A standard way to construct the metric from the ``zuvielbein'' $A-\bar A$ is by taking the trace of its square, $g_{\mu\nu}=\tfrac12\,\tr[(A-\bar A)_\mu(A-\bar A)_\nu]$. Plugging into this definition the asymptotic connections \eqref{AdSxR:BCs2} yields the asymptotic Lobachevsky line-element
\eq{
\extd s^2 = 2\tr(L_0)^2\,\extd\rho^2 + \extd t^2 + \big(\tfrac14\,e^{2\rho}+{\cal O}(1)\big)\,\extd\varphi^2 + {\cal O}(1)\,\extd t\extd\varphi\,.
}{eq:ds}

Following the algorithm outlined in \cite{Afshar:2012nk} with the boundary conditions above leads to the $W_N^{(2)}$ algebra discussed in section \ref{se:3.1} times another $\hat u(1)$ current algebra with level $\kappa$ \eqref{eq:hsg11} as quantum ASA. The current algebra part within the $W_N^{(2)}$ algebra reads
\eq{
[J_n,\,J_m]=-\tr(S^2)\,k_{\textrm{\tiny CS}}\,\delta_{n+m,\,0}\,.
}{eq:hsg33}
Comparing with \eqref{eq:FSa} relates the level $\hat u(1)$ level $\kappa$ with the Chern--Simons level $k_{\textrm{\tiny CS}}$.
\eq{
\kappa=-\tr(S^2)\,k_{\textrm{\tiny CS}}
}{kCSell}

In order to compute $\tr(S^2)$ we use the following definition of the Killing form, $K_{ab}=f^{d}{}_{ac}\,f^{c}{}_{bd}$, where $f^{d}{}_{ac}$ are the structure constants, and normalize it such that the metric of the $sl(2)$ part is
\eq{
\tr(L_aL_b)=\mathcal F (N)^{-1}K_{ab}=\left(\begin{matrix}
0  &  0&  -1\\
0  &  1/2&  0\\
-1  &  0&  0
   \end{matrix}\right)_{ab}\,.
}{eq:hsg34}
Using the general commutator with a field $\phi_\ell$ of weight $h$,
$[L_0,\,\phi_\ell]=-\ell\,\phi_\ell$,
yields
\eq{
\frac12\,\mathcal F (N)=2\!\!\sum_{\ell=-\tfrac{N}{2}+1}^{\tfrac{N}{2}-1}\!\!\ell^2+\sum_{h=2}^{N-1}\sum_{\ell=-h+1}^{h-1}\ell^2 = \frac{N^2(N-1)(N-2)}{6}
}{eq:lalapetz}
On the other hand, using the commutator
$[S,\,\psi_a^{\pm}]=\pm \psi_a^{\pm}$
we have $K_{SS}=2(N-1)$. So we conclude
$\tr(S^2)=\mathcal F (N)^{-1}K_{SS}=6/[N^2(N-2)]$.
The relation between $k_{\text{CS}}$ and $\alpha$ by virtue of \eqref{kCSell}, \eqref{eq:hsg11} and \eqref{eq:hsg20} finally yields
\eq{
k_{\textrm{\tiny CS}} = -\frac{N(N-2)}{6}\,\alpha = -\frac{N(N-2)}{N-\hat N-1}\,\frac{\hat N}{6}\,.
}{eq:hsg28}

Let us briefly comment on the sign in \eqref{eq:hsg28}: Since $\alpha\geq 0$ the Chern--Simons level is non-positive, which could easily be switched into non-negative by a parity flip or by exchanging $A\leftrightarrow\bar A$ in \eqref{eq:hs1}, so there is nothing profound about this sign. The signs that matter are the ones we checked in section \ref{se:3.3}.
The result \eqref{eq:hsg28} implies that the Chern--Simons level, whenever it is non-vanishing, is always large in the large $N$ limit. In the quantum regime it scales like $k_{\textrm{\tiny CS}}\propto N$, while in the semi-classical and the dual quantum regime it scales like $k_{\textrm{\tiny CS}}\propto N^2$. Therefore, even in the quantum regimes the saddle-point approximation in path-integral evaluations on the gravity side might be trustworthy. This is quite different from the situation encountered in \cite{Castro:2011zq}. 

%AdS boundary conditions can be extracted from \eqref{AdSxR:BCs2} as well, just by replacing the boundary conditions for $\bar A_\varphi$ by an expression analog to $A_\varphi$. This leads to two copies of the $W_N^{(2)}$ algebra discussed in section \ref{se:3.2} as quantum ASA.  Thus, it is possible to obtain unitary AdS holography at large central charge for the next-to-principal embedding.

AdS boundary conditions can be extracted from \eqref{AdSxR:BCs2} as well, just by replacing the boundary condition for $\bar A$ by an expression analog to $A$. This leads to two copies of the $W_N^{(2)}$ algebra discussed in section \ref{se:3.2} as quantum ASA.  Thus, it is possible to obtain unitary AdS holography at large central charge for the next-to-principal embedding.

\section{Conclusions}\label{se:4}

We have shown by explicit construction that 3-dimensional higher spin gravity with a specific non-principal embedding of $sl(2)$ into $sl(N)$ can lead to unitary models with arbitrarily large central charges for certain rational values of the Chern--Simons level $k_{\textrm{\tiny CS}}$ in \eqref{eq:hs1}. While for each given $N$ there is a unitarity bound  \eqref{eq:hsg4} on the central charge, we can obtain macroscopically large values of the central charge by allowing very large but finite values of $N$. In this sense, semi-classicality is not at odds with unitarity, thereby circumventing the no-go result of \cite{Castro:2012bc}.

An interesting feature that we observed is the behavior of the central charge $c$ as a function of the Chern--Simons level $k_{\textrm{\tiny CS}}$. The allowed values of $c-1$ \eqref{eq:hsg21} coincide precisely with the values of the central charge in the $W_{\hat N}$-minimal models, where $\hat N$ depends on $k_{\textrm{\tiny CS}}$ as given in \eqref{eq:hsg28}.
For the smallest values possible for $k_{\textrm{\tiny CS}}$ we are in a quantum regime with an almost integer central charge, $c=0,\,1,\,2-{\cal O}(1/N^2),\, \dots$ As we increase the Chern--Simons level the theory gets more and more semi-classical, eventually coming close to saturating the bound \eqref{eq:hsg4} for the central charge. In this semi-classical regime the allowed values of the Chern--Simons level are so close to each other that the central charge becomes a quasi-continuous parameter. As we keep increasing the Chern--Simons level, at some point the central charge turns around, as in figure \ref{fig:1}, and approaches ultimately a dual quantum regime, in which the level spacing of the central charge is $4+{\cal O}(1/N)$. The endpoint in the dual quantum regime is $c=1$ [$c=3-{\cal O}(1/N)$] if $N$ is odd [even].
In conclusion, 3-dimensional higher-spin gravity for non-principal embeddings can lead to novel unitary toy models for quantum gravity. 

We close with a list of open issues.
Our discussion was purely algebraical and focused on unitarity, with no attempt to consider partition functions or additional consistency requirements, like modular invariance. It would be interesting to generalize the discussion of \cite{Castro:2011zq} to $W_N^{(2)}$ gravity to verify how many of such models survive these checks. A key difference to their analysis is that our case leads to a discrete (but arbitrarily large) set of allowed values for the Chern--Simons level $k_{\textrm{\tiny CS}}$ and the central charge $c$. This is quite different from the situation in the principal embedding or Einstein gravity, where unitarity alone does not restrict the coupling constant in any way. A continuous coupling constant leads to a continuous central charge, and is thus at odds with the Zamolodchikov $c$-theorem \cite{Zamolodchikov:1986gt}. By contrast, even if there were no additional consistency requirements leading to further restrictions of the coupling constant, in our case there is no obvious violation of the Zamolodchikov $c$-theorem. 

At generic $N$ we considered exclusively the next-to-principal embedding. There are plenty of other non-principal embeddings. It could be interesting to classify this zoo and to establish which of them are compatible with unitarity and thus candidates for interesting quantum gravity models, or at least to find other infinite families of models compatible with semi-classical unitarity in the spirit of the present work.

It also seems likely that it is possible to construct a unitary theory by taking a scaling limit $N\rightarrow\infty$
with $\alpha$ fixed. If we restrict ourselves to considering only unitary representations of $W_N^{(2)}$, taking the large $N$ limit
of the $\mathcal{U}_q\big(sl(N|1)\big)$ construction results in a theory that matches the bosonic sector of the $\mathcal{N}=1$
supersymmetric extension of the $hs(\lambda)$ theory, with integer spins ranging from $1$ to $\infty$.

\section*{Acknowledgments}

We thank Per Kraus and Simon Ross for the kind invitation to contribute to a special issue of Classical and Quantum Gravity on ``Higher spins and holography''. 
% and Alejandra Castro for reading the manuscript.

HA, MG, DG and MR were supported by the START project Y~435-N16 of the Austrian Science Fund (FWF) and the FWF project I~1030-N27. 
Additionally, HA was supported by the FWF project I~952-N16 and DG by the FWF project P~21927-N16.
RR was supported by the FWF project P~22000-N16 and the project DO~02-257 of the Bulgarian National Science Foundation (NSFB).
This research was supported in part by the National Science Foundation under Grant No. NSF PHY11-25915.

\section*{References}

%\bibliography{review}

\providecommand{\href}[2]{#2}\begingroup\raggedright\endgroup

\end{document}